\documentclass[prb,aps,twocolumn,showpacs,floats]{revtex4}
\usepackage{epsfig,bm,epsf,graphics}
\begin{document}
%\draft
\title{Phase Transition in Heisenberg  Stacked Triangular Antiferromagnets: End of a Controversy}
\author{V. Thanh Ngo$^{a,b}$ and H. T. Diep\footnote{ Corresponding author, E-mail:diep@u-cergy.fr }}
\address{Laboratoire de Physique Th\'eorique et Mod\'elisation,
CNRS-Universit\'e de Cergy-Pontoise, UMR 8089\\
2, Avenue Adolphe Chauvin, 95302 Cergy-Pontoise Cedex, France\\
$^a$ Institute of Physics, P.O. Box 429,   Bo Ho, Hanoi 10000,
Vietnam\\
$^b$ Department of Physics, Tokyo Institute of Technology, 2-12-1 Ookayama, Meguro-ku, Tokyo
152-8551, Japan}

\begin{abstract}
By using the
Wang-Landau flat-histogram Monte Carlo (MC) method for very large lattice sizes never
simulated before, we show that the phase transition in the frustrated Heisenberg stacked triangular
antiferromagnet is of first-order, contrary to results of earlier MC simulations using old-fashioned
methods. Our result lends support to
the conclusion of a nonperturbative renormalization group performed on an
effective Hamiltonian. It puts an end to a 20-year long
controversial issue.
\end{abstract}
\pacs{75.10.-b  General theory and models of magnetic ordering ;
75.40.Mg    Numerical simulation studies}
\maketitle

\section{Introduction}

When a spin cannot fully satisfy energetically all the interactions with its neighbors, it is
"frustrated".   This situation occurs when the interactions are in competition with
each other or when the lattice geometry does not allow to satisfy all interaction bonds simultaneously
as seen for example in the triangular lattice  with an antiferromagnetic interaction between the
nearest-neighbors.
Effects of the frustration
in spin systems have been extensively
investigated during the last 30 years. Frustrated spin systems are shown to
have unusual properties such as large ground state (GS)
degeneracy, interesting GS symmetries, successive phase transitions
with complicated nature, partially disordered phase, reentrance and disorder lines.
Frustrated systems still constitute at present a challenge for
theoretical, experimental and simulational methods.  For  recent reviews, the
reader is referred to Ref. \onlinecite{Diep2005}.

The nature of the phase transition
in strongly frustrated spin systems has been a
subject of intensive investigations in the last 20 years.
Theoretically, these systems are excellent testing grounds for
theories and approximations.  Many well-established methods such as
renormalization group (RG), high- and low-temperature series
expansions etc often failed to deal with these systems.
Experimentally, data on different frustrated systems show a variety
of possibilities: first-order or second-order transitions with
unknown critical exponents etc. (see reviews in Ref.
\onlinecite{Diep2005}).  One of the most
studied systems is the stacked triangular antiferromagnet (STA): the
antiferromagnetic (AF) interaction between nearest-neighbor (NN)
spins on the triangular lattice causes  a very strong frustration.
It is impossible\cite{Diep2005} to fully satisfy the
three AF bond interactions on each equilateral triangle.  The GS
configuration of both Heisenberg and XY models is the well-known
120-degree structure.
The cases of XY ($N=2$) and Heisenberg ($N=3$) spins on the
STA have been intensively studied since 1987.  For details, see for example the review by Delamotte
et al\cite{Delamotte2004}.
Let us briefly recall here some main
historical developments.
Kawamura~\cite{kawamura87,kawamura88} has  conjectured by a
two-loop RG analysis  and Monte Carlo (MC) simulations that the
transition in $XY$ and Heisenberg models belong each to
a new universality class in dimension $d=3$. Since then there have been many other
calculations and simulations with contradictory
results.  For example, Azaria et al\cite{azaria90}
suggested from a non-linear sigma model
that if the transition is  not of first order or
mean-field tricritical then it should be $O(4)$ universality.
Numerical
simulations\cite{Loison,Boub,Dobry} however did not confirm these conjectures.
Antonenko et al.\cite{antonenko95} went further in a four-loop RG calculation with a Borel
resummation technique. They
concluded that the transition is of first order.
From 2000,  Tissier and coworkers
~\cite{tissier00b,tissier00,tissier01} have carried out a nonperturbative RG
study of frustrated magnets for any dimension between  two and four.
They recovered all known perturbative one-loop results in two and
four dimensions as well as for the infinite spin-component number
 $N\to \infty$. They determined
$N_c(d)$ for all $d$ and found $N_c(d=3)=5.1$ below which the
transition is of first order  in contradiction with the
conjecture of the existence of a
new chiral universality class by Kawamura.\cite{kawamura87,kawamura88}
They explained why theories and simulations have encountered so far many difficulties by
the existence of a
whole region in the flow diagram in which the flow is slow: the first-order character for $N=2,3$ is so weak that the
transition has a second-order aspect with "pseudo" critical exponents.  They
calculated these pseudo exponents and found that
they coincided with some experimental data.  While this scenario is very coherent, we note that
in this nonperturbative RG
technique, the real Hamiltonian is truncated at the beginning and replaced by
an effective one.   However, as
will be seen in this paper, the nonperturbative results are
well confirmed.

Let us recall some results on the XY case.  Early MC results on XY STA have been reviewed by
Loison.\cite{Loison}  Until 2003, all numerical simulations found
a second-order transition with exponents.
A numerical breakthrough has been realized with the
results of Itakura\cite{itakura03} who used an improved MC
renormalization-group scheme to investigate the
renormalization group flow of the
   effective Hamiltonian used in
  field-theoretical studies for the XY STA. He found that the XY STA
  exhibits a clear first-order behavior and there are no chiral
  fixed points of renormalization-group
  flow for $N$=2.
In 2004, Peles et al\cite{Peles} have used a continuous model to
study the XY STA by MC simulation. They found evidence of a
first-order transition.  In 2006, Kanki et al\cite{Kanki}, using a
microcanonical MC method, have found a first-order signature of the
XY STA.
While these recent simulations have demonstrated evidence of
first-order transition for the XY STA in agreement with the
nonperturbative RG analysis, all of them suffered one or two uncertain
aspects: the work of Itakura has used a truncated Hamiltonian, the
work of Peles et al has used standard MC methods and the work of
Kanki et al used a traditional microcanonical MC technique.
Using a very high-performance technique for weak
first-order transitions, the so-called Wang-Landau flat-histogram method,\cite{WL1}
we have recently carried out  simulations on the XY STA. We have found clearly
a first-order transition in that system confirming results of other
authors and putting an end to the
controversy which has been lasting for 20 years.

For the Heisenberg case,
Itakura\cite{itakura03} found, as in the XY case mentioned above, the absence of chiral
fixed points of renormalization-group
flow.   However, he could not find  numerical evidence of the first-order transition.
He predicted that if the transition is of first order for the Heisenberg spins,
it should occur at much
larger lattice sizes which he was not able to perform at that time.
Encouraged by the
high performance of the Wang-Landau method, we decided to study the Heisenberg case
in this work using the full Hamiltonian with very large lattice sizes.
As shown below, we find indeed a first-order transition in this case.

The paper is organized as follows. Section II is devoted to the
description of the model and  the technical details of the Wang-Landau
(WL) methods as applied in the present paper.  Section III shows our
results.  Concluding remarks are given in section IV.

\section{Monte Carlo Simulation: Wang-Landau algorithm}

We consider the stacking of triangular lattices in the $z$ direction. The spins are
the classical Heisenberg model of magnitude
$S=1$. The Hamiltonian is given by
\begin{equation}
{\cal H} = J\sum_{(i,j)} \mathbf{S}_i.\mathbf{S}_j,
\end{equation}
where $S_i$ is the Heisenberg spin at the lattice site $i$, $\sum_{(i,j)}$ indicates the
sum over the NN spin pairs $S_i$ and $S_j$ both in the $xy$ planes and in adjacent planes in the $z$ direction.
For simplicity, we suppose the same antiferromagnetic interaction $J$ ($J>0$) for both in-plane NN pairs and
inter-plane NN ones.

Recently, Wang and Landau\cite{WL1} proposed a Monte Carlo algorithm for classical statistical models. The algorithm uses a random walk in energy space in order to obtained an accurate estimate for the density of states $g(E)$ which is defined as the number of spin configurations for any given $E$. This method is based on the fact that a flat energy histogram $H(E)$ is produced if the probability for the transition to a state of energy $E$ is proportional to $g(E)^{-1}$.
At the beginning of the simulation, the density of states (DOS) is set equal to one for all possible energies, $g(E)=1$.
We begin a random walk in energy space $(E)$ by choosing a site randomly and flipping its spin with a probability
proportional to the inverse of the momentary density of states. In general, if $E$ and $E'$ are the energies before and after a spin is flipped, the transition probability from $E$ to $E'$ is
\begin{equation}
p(E\rightarrow E')=\min\left[g(E)/g(E'),1\right].
\label{eq:wlprob}
\end{equation}
Each time an energy level $E$ is visited, the DOS is modified by a modification factor $f>0$ whether the spin flipped or not, i.e. $g(E)\rightarrow g(E)f$.
  At the beginning of the random walk, the modification factor $f$ can be as large as $e^1\simeq 2.7182818$. A histogram $H(E)$ records how often a state of energy $E$ is visited. Each time the energy histogram satisfies a certain "flatness" criterion, $f$ is reduced according to $f\rightarrow \sqrt{f}$ and $H(E)$ is reset to zero for all energies. The reduction process of the modification factor $f$ is repeated several times until a final value $f_{\mathrm{final}}$ which close enough to one. The histogram is considered as flat if
\begin{equation}
H(E)\ge x\%.\langle H(E)\rangle
\label{eq:wlflat}
\end{equation}
for all energies, where $x\%$ is chosen between $70\%$ and $95\%$
and $\langle H(E)\rangle$ is the average histogram.

The thermodynamic quantities\cite{WL1,brown} can be evaluated by
%\begin{eqnarray}
$\langle E^n\rangle =\frac{1}{Z}\sum_E E^n g(E)\exp(-E/k_BT)$,
$C_v=\frac{\langle E^2\rangle-\langle E\rangle^2}{k_BT^2}$,
$\langle M^n\rangle =\frac{1}{Z}\sum_E M^n g(E)\exp(-E/k_BT)$, and
$\chi=\frac{\langle M^2\rangle-\langle M\rangle^2}{k_BT}$,
%\end{eqnarray}
where $Z$ is the partition function defined by
%\begin{equation}
$Z =\sum_E g(E)\exp(-E/k_BT)$.
%\label{eq:partfunc}
%\end{equation}
The canonical distribution at any temperature can be calculated simply by
%\begin{equation}
$P(E,T) =\frac{1}{Z}g(E)\exp(-E/k_BT)$.
%\label{eq:pe}
%\end{equation}

In this work, we consider a energy range of interest\cite{Schulz,Malakis}
$(E_{\min},E_{\max})$. We divide this energy range to $R$ subintervals, the minimum energy of each subinterval is $E^i_{\min}$ for $i=1,2,...,R$, and maximum of the subinterval $i$ is $E^i_{\max}=E^{i+1}_{\min}+2\Delta E$,
where $\Delta E$ can be chosen large enough for a smooth boundary between two subintervals. The Wang-Landau
algorithm is used to calculate the relative DOS of each subinterval $(E^i_{\min},E^i_{\max})$ with the
modification factor $f_\mathrm{final}=\exp(10^{-9})$ and flatness criterion $x\%=95\%$.
We reject the suggested spin flip and do not update $g(E)$ and the energy histogram $H(E)$ of
the current energy level $E$ if the spin-flip trial would result in an energy outside the energy segment.
The DOS of the whole range is obtained by joining the DOS of each
subinterval $(E^i_{\min}+\Delta E,E^i_{\max}-\Delta E)$.

\section{Results}

 We used the system size of $N\times N \times N$ where
$N=72$, 84, 90, 96, 108, 120 and 150. Periodic
boundary conditions are used in the three directions.  $J=1$ is
taken as the unit of energy in the following.

\begin{figure}
\centerline{\epsfig{file=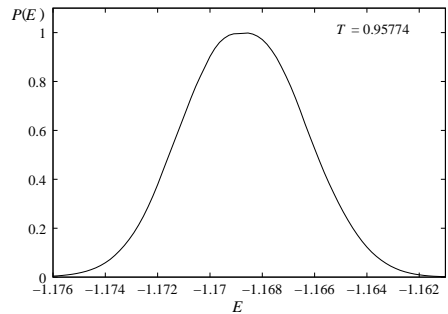,width=2.3in}} \caption{Energy
histogram for $N=96$ at $T_c$ indicated on the
figure.}\label{fig:STAH96PE}
\end{figure}

\begin{figure}
\centerline{\epsfig{file=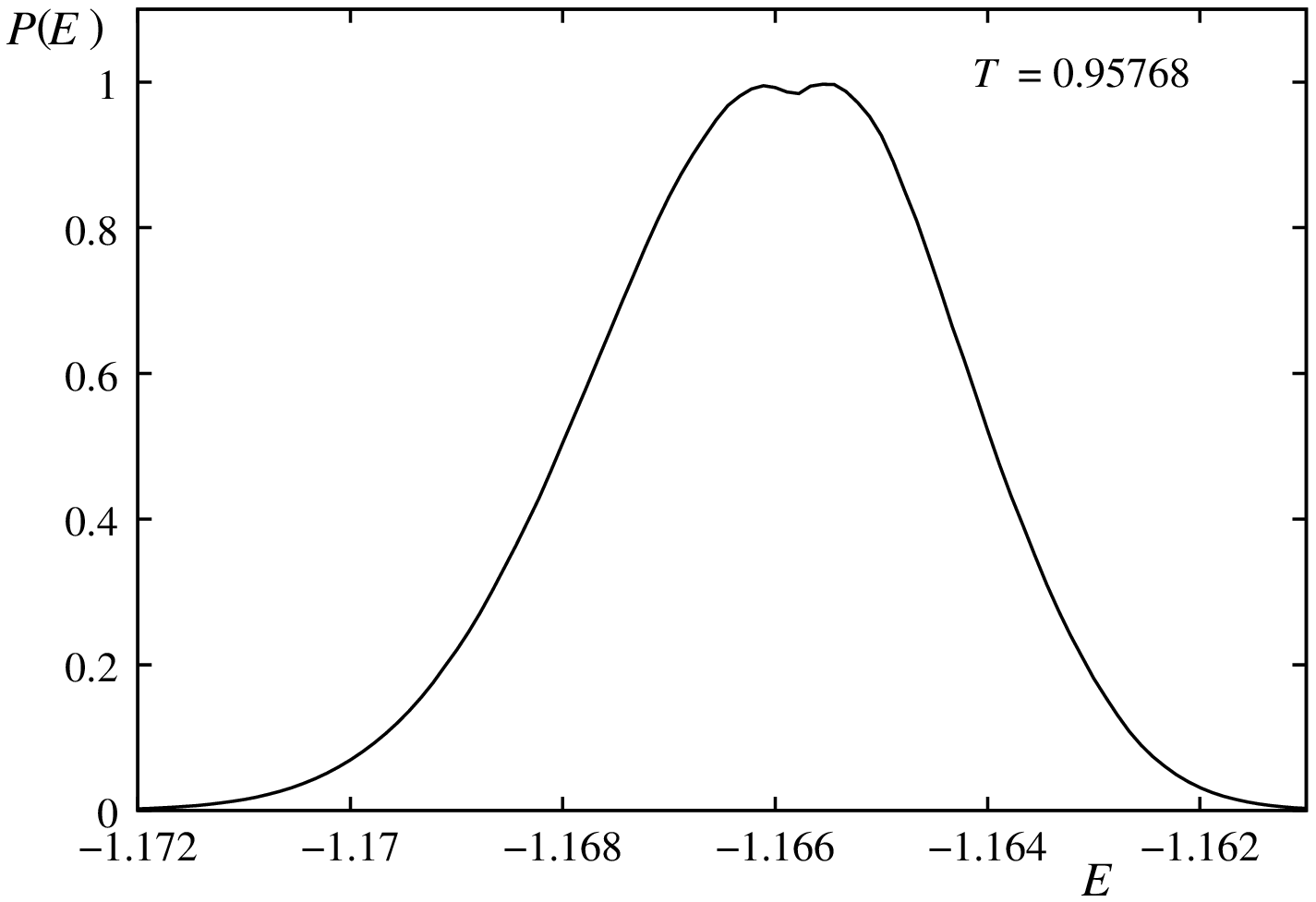,width=2.3in}} \caption{Energy
histogram for  $N=120$ at $T_c$ indicated on the figure.}
\label{fig:STAH120PE}
\end{figure}

\begin{figure}
\centerline{\epsfig{file=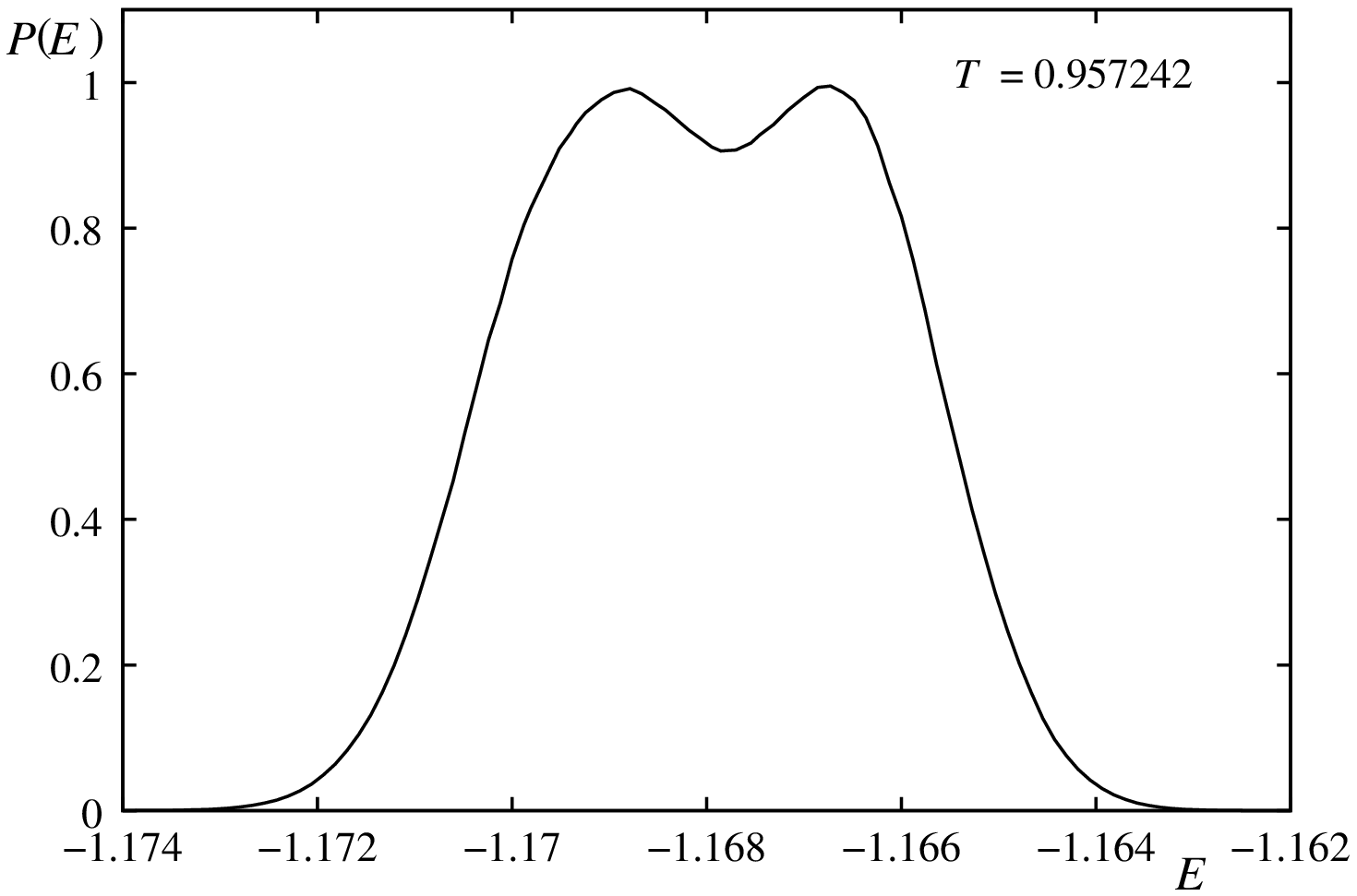,width=2.3in}} \caption{Energy histogram
for $N=150$ at $T_c$ indicated on the figure.} \label{fig:STAH150PE}
\end{figure}

The energy histograms for three representative sizes $N=96$, $N=120$ and
$N=150$ shown in Figs. \ref{fig:STAH96PE},  \ref{fig:STAH120PE} and \ref{fig:STAH150PE},
respectively.  As seen, for $N=96$, the peak is very broad, a signature of the beginning to
of a double-maximum structure. The double peak begins really at $N=120$.  We note
that the distance between the two peaks, i. e.
the latent heat, increases with increasing size and reaches $0.0025$ for $N=150$.  This is to be compared
with the value $\simeq 0.009$ for $N=120$ in the XY case.\cite{itakura03,Peles,Kanki,Ngo08}
Such a small value of the latent heat in the Heisenberg case explains why the
first-order character was so difficult to be observed. 
For increasing sizes, the
minimum between the peaks will be deepened to separate completely the two peaks.  
Note that the double-peak structure is a sufficient
condition, not a necessary condition, for a first-order transition.
We give here the values of $T_c$ for a few sizes:  $T_c =$
0.95774, 0.95768 and 0.957242 for $N$=96, 120 and 150,
respectively.

\begin{figure}
\centerline{\epsfig{file=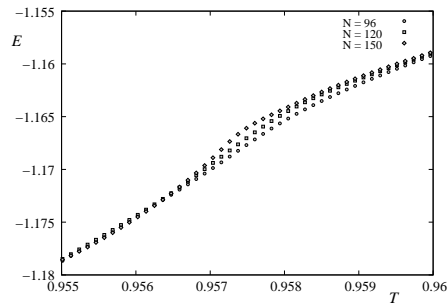,width=2.3in}} \caption{Energy
versus $T$ for $N=96$, 120, 150.}\label{fig:STAH3E}
\end{figure}

To explain why standard MC methods  without
histogram monitoring (see for example Ref. \onlinecite{kawamura87})
fail to see the first order character, let us show in Fig.
\ref{fig:STAH3E} the energy vs $T$ obtained by averaging over
states obtained by the WL method for $N=96$, 120 and 150. We see here that while
the energy histograms show already a signature of double-peak structure
at these big sizes, the average energy calculated by using these WL histograms
does not show a discontinuity: the averaging over all states
erases away the bimodal distribution seen  in the energy histogram
at the transition temperature.  Therefore, care should be taken to
avoid such problems due to averaging in MC simulations when studying weak first-order transitions.

%\begin{figure}
%\centerline{\epsfig{file=STA36PE.eps,width=2.8in}} \caption{Energy
%histogram for $N=36$.} \label{fig:STA36PE}
%\end{figure}

%\begin{figure}
%\centerline{\epsfig{file=STA36GE.eps,width=2.8in}}
%\caption{Density of state for $N=36$.} \label{fig:STA36GE}
%\end{figure}


%\begin{figure}
%\centerline{\epsfig{file=STAH3CV.eps,width=2.5in}} \caption{Specific heat
% versus $T$  for  $N=96, 120, 150$.}
%\label{fig:STAH3CV}
%\end{figure}

\begin{figure}
\centerline{\epsfig{file=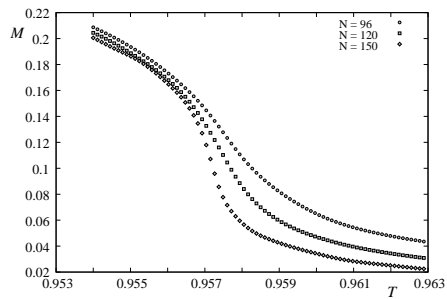,width=2.3in}} \caption{Magnetization
 versus $T$ for $N=$96, 120, 150.} \label{fig:STAH3M}
\end{figure}

\begin{figure}
\centerline{\epsfig{file=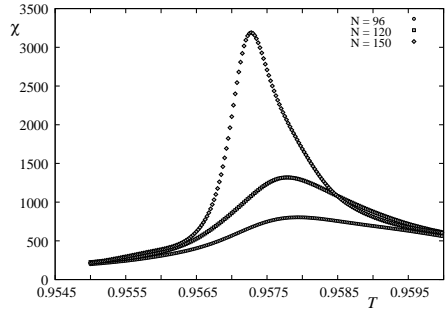,width=2.3in}} \caption{Susceptibility
 versus $T$ for $N=$96, 120, 150.} \label{fig:STAH3X}
\end{figure}

Figures \ref{fig:STAH3M} and \ref{fig:STAH3X} show the magnetization and the susceptibility
for three sizes $N=96$, 120 and 150.  Again here, one does not see with one's eye the discontinuity
of the magnetization at the transition
even for $N=150$. The averaging procedure erases, as for the energy, the detailed structure
at the transition.

\begin{figure}
\centerline{\epsfig{file=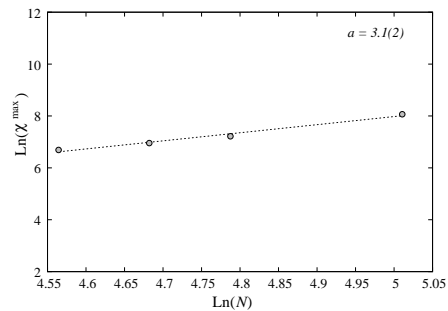,width=2.3in}} \caption{Maximum of susceptibility
 versus $N$=96, 108, 120 and 150 in the $\ln-\ln$ scale. The slope is 3.1. 
 See text for comments.} \label{fig:STAHGAM}
\end{figure}

At this stage it is interesting to make another check of the first-order character: in a
first-order transition, the maximum of the susceptibility should scale with the system volume,
namely $N^d$ where $d$ is the system dimension.\cite{Challa} We plot in Fig. \ref{fig:STAHGAM} $\chi^{max}$ versus $N$
in a $\ln-\ln$ scale.  The slope of the straight line is $\sim 3.1$ which is nothing but $d$ within errors.  This
is a very strong signature of a first-order transition.

\section{Concluding Remarks}

We have studied in this paper the phase transition in the Heisenberg STA by
using the flat histogram technique invented by Wang and Landau. The
method is very efficient because it helps to overcome extremely long
transition time between energy valleys in systems with a
first-order phase transition.   We found that the transition becomes
clearly of first-order only at a very large lattice size confirming
the result of a nonperturbative RG
calculations using an effective average Hamiltonian. The present work hence
puts definitely an end to the long-standing
controversial subject on the nature of the phase transition in Heisenberg STA.
To conclude, let us emphasize
that for complicated systems like this one, methods well established for simple systems
such as ferromagnets may encounter difficulties in dealing with the nature of the phase transition.
Such difficulties can be solved only with high-performance MC simulations as the one used here, and a detailed analysis of the flow behavior as suggested by a nonperturbative RG calculation.

One of us (VTN) would like to thank  Nishina Memorial Foundation for a six-month
postdoctoral fellowship. He is also grateful to Prof. T. Ando for hospitality
and encouragement during his stay at the
Tokyo Institute  of Technology.

{}

\end{document}